\documentclass[apj]{emulateapj}

\usepackage{aas_macros}
\usepackage{apjfonts}
\usepackage{graphicx}
\usepackage{color}
\usepackage{amsbsy}
\usepackage{mathrsfs}
\usepackage{mathpazo,bm}
\usepackage{amsmath}
\usepackage{multirow}
\usepackage{epstopdf}
\usepackage{amssymb}
\usepackage{natbib}

\newlength{\hfwidth}
\newlength{\hfwidthsingle}
\renewcommand{\v}[1]{{\boldsymbol{#1}}} 

\newcommand{\del}{\v{\nabla}}

\newcommand{\Eq}[1]{Eq. (\ref{#1})}

\newcommand{\eq}[1]{\Eq{#1}}

\newcommand{\Fig}[1]{Fig.~\ref{#1}}
\newcommand{\fig}[1]{\Fig{#1}}

\newcommand{\BV}{Brunt-V\"ais\"al\"a frequency}

\definecolor{brown}{rgb}{0.42,0.24,0.07}
\definecolor{darkgreen}{rgb}{0.0,0.6,0.00}
\definecolor{purple}{rgb}{0.7,0.0,0.7}
\definecolor{black}{rgb}{0.0,0.0,0.0}

\shorttitle{A Parameter Study for Baroclinic Vortex Amplification}
\shortauthors{Raettig et al.}

\begin{document}


\title{A Parameter Study for Baroclinic Vortex Amplification}


\author{Natalie Raettig\altaffilmark{1,2}
        Wladimir Lyra\altaffilmark{2,3,4}, and
      Hubert Klahr\altaffilmark{1}
    }

\altaffiltext{1}{Max-Planck-Institut f\"ur Astronomie, K\"onigstuhl 
                 17, 69117, Heidelberg, Germany;
                 raettig@mpia.de,klahr@mpia.de}
\altaffiltext{2}{Department of Astrophysics, American Museum of
                 Natural History, 79th Street at Central Park 
                 West, New York, NY,10024, USA}
\altaffiltext{3}{Jet Propulsion Laboratory, California Institute of Technology, 4800 Oak Grove Drive, Pasadena, CA 91109, USA; Wladimir.Lyra@jpl.nasa.gov}
\altaffiltext{4}{NASA Carl Sagan Fellow}




\begin{abstract}



Recent studies have shown that baroclinic vortex amplification is strongly dependent on certain factors, namely, the global entropy gradient, the efficiency of thermal diffusion and/or relaxation as well as numerical resolution. We conduct a comprehensive study of a broad range and combination of various entropy gradients, thermal diffusion and thermal relaxation time-scales via local shearing sheet simulations covering the parameter space relevant for protoplanetary disks. We measure the Reynolds stresses as a function of our control parameters and see that there is angular momentum transport even for entropy gradients as low as $\beta=-{d\ln s}/{d\ln r}={1}/{2}$. Values we expect in protoplanetary disks are between $\beta=0.5-2.0$ The amplification-rate of the perturbations, $\Gamma$, appears to be proportional to $\beta^2$ and thus proportional to the square of the \BV ($\Gamma \propto \beta^2 \propto N^2$). The saturation level of Reynolds stresses on the other hand seems to be proportional to $\beta^{1/2}$. This highlights the importance of baroclinic effects even for the low entropy gradients expected in protoplanetary disks. 
\end{abstract}


\keywords{accretion, accretion disks, circumstellar matter, hydrodynamics, instabilities, turbulence, methods: numerical, solar system: formation, planetary systems }

\section{Introduction}

Angular-momentum transport and turbulence are important issues concerning protoplanetary disks. Magneto-hydrodynamic turbulence brought about by the magnetorotational instability  (MRI, \citealt{BalbusHawley1991}), is a reliable way to achieve a sufficient angular-momentum transport and with this also an accretion rate fitting observations \citep{Andrewsetal2009} and playing an important role in planet formation \citep{ Johansenetal2007, Lyraetal2008, Nataliaetal2010, Flocketal2011,Uribe2011, Johansen2011}. However, for MRI to be active the gas has to be sufficiently ionized. This is only the case in the outer regions, upper layers of the disk, and in regions close to the star. The other parts of the disk are too cold and dust-rich for sufficient ionization and thus the magnetic fields cannot couple to the gas. Because of this, the MRI cannot operate in this region, which is therefore called "dead zone" \citep{Gammie1996,TurnerDrake2009}.

Since the precise ionization structure is still under debate \citep{TurnerDrake2009} as is the interplay between active and dead-zones \citep{LyraMacLow2012} we want to assess the precise hydrodynamic behavior of dead zones, because accretion has to proceed through it somehow and it is where planets form. Therefore it is of interest to study purely hydrodynamic turbulence in circumstellar disks. \citet{KlahrBodenheimer2003} found such a hydrodynamic instability creating vortices in three-dimensional radiation hydrodynamical simulations of baroclinic disks, e.g.\ with a radial entropy gradient and thus vertical shear, which they assumed to be a kind of baroclinic instability (BI) modified by the Keplerian shear profile. Observed protoplanetary disks have a non-zero radial entropy gradient $\beta=-{d\ln s}/{d\ln r}$, where $s$ is the entropy and $r$ the radial distance to the star. With $\beta=q-\left(\gamma_{\rm2D}-1\right)p_\Sigma$, where $q=-{d\ln T}/{d\ln r}$ and  $p_\Sigma=-{d\ln \Sigma}/{d\ln r}$ are the temperature surface density gradient respectively and $\gamma_{\rm2D}$ is the 2D adiabatic index, we see that disks that fulfill $p_\Sigma<q/(\gamma_{\rm 2D}-1)$ indeed have a negative entropy gradient with values from \citet{Andrewsetal2009} of $q\approx 0.3-0.5$ and $p_\Sigma=0.9$. Therefore protoplanetary disks are not barotropic but rather baroclinic which means that planes of constant pressure and constant density are misaligned, creating a thermal wind, e.g.\ vertical shear. In a linear stability analysis that followed \citep{Klahr2004} it was shown that this instability can only be of non-linear nature \citep[see also][]{Cabot1984, KnoblochSpruit1986}.

Thermal relaxation turned out to be crucial when \citet{PetersenJulienStewart2007, PetersenStewartJulien2007} studied baroclinic vortex amplification using an incompressible approximation. In fact thermal relaxation or diffusion, besides the entropy gradient, are key ingredient to establish baroclinic feedback that keeps the instability e.g. vortices in baroclinic disks growing. 

While both effects e.g. the baroclinic instability and baroclinic vortex amplification are a result of the superadiabatic radial stratification of a disk they are not to be confused. An operating linear baroclinic instability \citep[compare][]{Cabot1984, KnoblochSpruit1986} would be able to create vortices in disks from infinitesimal perturbations, whereas the baroclinic vortex amplification deals with the growth of existing vortical perturbations, for which \citet{LesurPapaloizou2010} used the term "subcritical baroclinic instability" (SBI).

The occurrence of a classical BI in the disk in its geophysical definition is still under debate and shall be discussed elsewhere. There are three possibilities: 1.) there is a classical BI working in protoplanetary disks creating the initial vortices, 2.) there is an other instability operating \citep[see the discussion in ][]{Klahr2004} for instance creating vortices via Kelvin-Helmholz instability from vorticity maxima in sheared waves of baroclinic disks or 3.) small vortical perturbations are triggered from other effects, e.g. waves from the MHD active region of the disks or maybe from the waves emitted by vortices at other radii. In any case the vortices are then growing as described by the BVA until they reach a sufficient size to influence the evolution of the disk, and this is the physics being subject of the present paper.

Recently, \citet{LyraKlahr2011} have examined the interplay of baroclinic vortex amplification and MHD. They found that as soon as magnetic fields are coupled to the gas, the MRI takes over and thus superseeds vortices which were previously amplified by vortex amplification. This is evidence that the vortex amplification is a phenomenon restricted to the dead-zone. 

All the above mentioned (lower resolution) studies had to apply entropy gradients 2-4 times stronger than to be expected in protoplanetary disks \citep[][Klahr 2013 submitted]{Andrewsetal2009} to drive BVA. We show in the current  paper, through high resolution runs that realistic entropy gradients in protoplanetary  disks are sufficient for BVA.

Recently \citet{Paardekooperetal2010} have investigated the effect of radial vortex migration. They discovered that vortices migrate quickly radially inward once grown to their full size. While this effect will be of major importance to understand the life-cylce of a vortex, it plays a weaker role for the small/still growing vortices in the present paper. Of course migration will influence the effective angular momentum transport generated by the vortices via the emission of waves, but this is beyond the scope of 2D local simulations as in our study. We shall return to vortex migration and have a better estimate for angular momentum transport once we return to global simulations.

We carry out local, compressible shearing sheet simulations at various resolutions. We show that as we go to higher resolutions one can excite the nonlinear instability and achieve Reynolds stresses with the low entropy gradients deduced for observed accretion disks. We conduct an extensive parameter study for entropy gradients ($\beta$), resolution, thermal cooling ($\tau_\mathrm{cool}$) and diffusion times ($\tau_\mathrm{diff}$) respectively. Section 2 gives a brief overview of the physical background of the instability. In Section 3 we present the numerical setup of our simulations. In Section 4 we examine the amplification and decay-times of values such as enstrophy $\omega_z^2=(\del\times\boldsymbol{u})_z^2$ and $\alpha$-stresses. Here $\alpha=\langle\rho u_xu_y(q p_0)^{-1}\rangle$ with $\rho$ being the gas density, $\boldsymbol u$ the gas velocity, $q=1.5$ the shear parameter, and $p_0$ the initial mean pressure. We also analyze the saturation values, e.g. how quantities like the entropy gradient, cooling processes in the disk or the size of the simulated domain influence the strength of angular momentum transport. Finally we summarize our results and give a conclusion in Section 5.

\section{Physical Background}\label{PB}

Vorticity is conserved in quasi-incompressible barotropic simulations, but in flows with density and pressure as independent quantities vorticity is produced via the so called baroclinic term
\begin{equation}
\frac{\partial \v \omega}{\partial t}=\nabla\times\left( -\frac{1}{\rho}\nabla p\right) = \frac{1}{\rho^2}\nabla\rho\times\nabla p \propto \beta\partial_y\rho.
\label{BI_term}
\end{equation}
\noindent Here $\rho$ is the gas density, $p$ the gas pressure, and $\beta$ is the global radial entropy gradient. The ground state of a disk is geostrophic, e.g. all centrifugal forces and gravity are in balance with the strictly radial pressure gradient. If an entropy perturbation is introduced without perturbing the pressure, then this entropy perturbation will efficiently create vorticity in the presence of the global entropy and pressure gradients. This effect is basically radial buoyancy because of superadiabatic radial stratification\footnote{Note that similar situations can be found in subadiabatic configurations. In fact, in any non barotropic disk, an entropy perturbation will lead to a vorticity fluctuation. But without the global pressure and entropy gradient pointing in the same direction these perturbations will quickly decay (shear away) as they are lacking the mechanism of vortex amplification.}. Indeed the radial \BV \ \citep{Tassoul2000}
\begin{equation}
N^2= -\frac{1}{\gamma\rho}\frac{\partial p}{\partial r}\frac{\partial}{\partial r}\ln\left(\frac{p}{\rho^\gamma}\right)
\end{equation}
\noindent is imaginary, which would lead to radial convection. However, shear stabilizes non-axisymmetric modes and for the dynamic stability of the axisymmetric system the Solberg-H\o iland criterium \citep{Tassoul2000, Rudigeretal2002}
\begin{eqnarray}
\label{SH}
\frac{1}{R^3}\frac{\partial j^2}{\partial R}-\frac{1}{C_p\rho}\nabla p \nabla S>0\\
\frac{\partial p}{\partial z}\left(\frac{\partial j^2}{\partial R}\frac{\partial s}{\partial z}-\frac{\partial j^2}{\partial z}\frac{\partial s}{\partial R}\right)<0\nonumber
\end{eqnarray}
has to be considered. If one re-writes \Eq{SH} for local approximation \citep[see e.g.][]{BalbusHawley1998} the stabilizing action of the specific angular momentum shows up as the value of Oort's constant in the Coriolis term.
If also the vertical stratification in velocity is taken under consideration, as it will occur in real three-dimensional accretion disks \citep{FromangLyra2011}, then the combined action of radial buoyancy and Coriolis forces lead to a thermal wind, e.g. a vertical shear in rotational velocity. This is precisely the initial state as baroclinic instability in rotating stars and planetary atmospheres. Yet, instability in these systems is not obstructed by radial shear, whereas in a Keplerian disk radial scales would have to be on the order of the vertical pressure scale-heigth ($H$) \citep{KnoblochSpruit1986} to be linearly unstable with respect to baroclinic instability.

Before we explain the motion of a gas parcel in a vortex we want to explain the cooling and heating processes in a disk as they proved to be crucial to maintain the baroclinic feedback \citep{PetersenJulienStewart2007, PetersenStewartJulien2007}. Dust particles absorb photons which heats them up. To cool they radiate photons in the infrared. This radiation can be absorbed by other particles. This happens on a typical length-scale. A convenient parametrization for the diffusion time in our vortex system is $\tau_{\rm diff}=a^2/K$ where $a$ is the radius of the vortex and $K$ the diffusion constant. The diffusion constant can be approached using a flux limited diffusion approach as in \citet{KleyBitschKlahr2009}. There $K=\lambda c4a_{\rm R}T^3\left(\rho\kappa\right)^{-1}$ where $\lambda$ is the flux limiter, $c$ the speed of light, $a_{\rm R}$ the radiation constant, $T$ and $\rho$ the gas temperature and density, respectively and $\kappa$ the opacity.  Since $K$ is constant and the vortex grows $\tau_{\rm diff}$ will change over time. Thermal relaxation is the other process by which dust can deposit heat into the gas. When a dust particle has a certain temperature other than the equilibrium temperature it will exchange heat with the ambient medium until it reaches the background temperature again. $\tau_{\rm cool}$ is the time needed to achieve this. This time-scale affects vortices of all sizes equally.

The baroclinic feedback itself was explained in detail by \citet{PetersenStewartJulien2007}. A nice description of the mechanism can also be found in \citet{LesurPapaloizou2010}. In a baroclinic flow entropy is a function of pressure and density, $s\left(p, \rho\right)$. Pressure on the other hand is only a function of radius. The vortex interior transports high entropy material from small radii to large radii. After thermalization low entropy material is transported to small radii. Since the pressure variations, especially from weak vortices, are negligible in comparison to the global radial pressure gradient and much smaller than the azimuthal entropy gradient, pressure can be seen as approximately azimuthally constant \citep{KlahrBodenheimer2003, Klahr2004, PetersenJulienStewart2007}. To keep the pressure constant an azimuthal density gradient is established,  e.g. outflowing material has a lower density as inflowing material. Thus the vortex feels the effect of differential buoyancy which established the positive baroclinic feedback (\Eq{BI_term}). 

If cooling is too fast (short time-scales) then the fluid parcel adapts the background temperature slope too quickly. The vortex becomes locally isothermal and no entropy transport is possible. Conversely,  if cooling is too slow (long time-scales) then gas will not be thermalized fast enough. The vortex gas becomes adiabatic with constant entropy across the vortex. In both extreme cases, isothermal or adiabatic, the azimuthal entropy gradient across the vortex vanishes. As shown in \eq{BI_term} the vorticity source ceases to amplify the vortex, or at least stabilizes it against losses from numerical viscosity from radiating vorticity perturbations, e.g. Rossby waves. Therefore it is important that thermal cooling and diffusion times are in the right regime.

We model both thermal relaxation and thermal diffusion separately because, dependent on the vortex size, either one or the other dominates thermalization. Always the process with the shorter time-scale sets the heat exchange between vortex and ambient gas.

\section{Numerical Setup}

Our simulations were conducted with the {\sc Pencil Code}\footnote{See http://www.nordita.org/software/pencil-code/}. We use a two-dimensional, local shearing sheet approach. We consider a sheet in the mid-plane that co-rotates with the co-rotational radius $R_0$. This is  a 2D version of the model used in \citet{LyraKlahr2011}. To include the baroclinic term they define a global entropy gradient $\beta$. Note that in our approximation the gradients for entropy ($s$) and pressure ($p$) are the same. Therefore we do not distinguish between them in our notation and call both $\beta$. However, in real disks both may easily differ.

The total pressure $p_{\rm tot}=\bar p + p$ consist of a local fluctuation $p$ and a time-independent part that follows a large scale radial pressure gradient $\beta$
\begin{equation}
\bar p=p_0(r/R_0)^{-\beta},
\end{equation}
\noindent where $r$ is the cylindrical radius. The full set of linearized equations used in our simulations is
\begin{eqnarray}
\frac{\mathcal{D}\rho}{\mathcal{D}t} &+&\left(\v{u} \cdot\nabla\right)\rho=-\rho\nabla\cdot\v{u}+ f_D(\rho)\\
\label{NS}\frac{\mathcal{D}\v{u}}{\mathcal{D}t}&+&\left(\v{u} \cdot\nabla\right)\v{u}=-\frac{1}{\rho}\nabla p - 2\Omega_0\left(\v{\hat{z}}\times\v{u}\right)\nonumber\\
&+& \frac{3}{2}\Omega_0u_x\v{\hat{y}}+\frac{\beta p_0}{R_0}\left(\frac{1}{\rho}-\frac{1}{\rho_0}\right)\v{\hat{x}}+f_\nu(\v{u},\rho)\\
\frac{\mathcal{D}{s}}{\mathcal{D}t}&+&\left(\v{u} \cdot\nabla\right){s}=\frac{1}{\rho T} \bigg\{\nabla \cdot \left(K\nabla T\right)-\rho c_v\frac{(T-T_0)}{\tau_{\rm cool}}    \nonumber\\
&+& \frac{\beta p_0}{R_0}\frac{u_x}{\left(\gamma -1\right)}\bigg\}+f_K(s).
\end{eqnarray}
Here $\rho$ is the gas density, $\v{u}$ is the deviation of the gas velocity from the Keplerian value, $T$ the temperature, $c_v$ the specific heat at constant volume and, $K$ the heat conductivity. Tthermal diffusion time-scale is denoted by $\tau_{\rm cool}$. The symbol
\begin{equation}
\frac{\mathcal{D}}{\mathcal{D}t}= \frac{\partial}{\partial t}+u_y^{\left(0\right)}\frac{\partial}{\partial y}
\end{equation}
\noindent represents the Keplerian derivative where $u_y^{\left(0\right)}=-3/2\Omega_0x$. 

For a more thorough derivation of these equations and the linearization of the global pressure gradient we refer to \citet{LyraKlahr2011} and the appendix therein.

In order to keep the numerical scheme stable we add sixth-order hyperdiffusion  $f_D(\rho)$, hyperviscosity $f_\nu(\v{u},\rho)$, and hyperconductivity $f_K(s)$ \citep{Lyraetal2008, Lyraetal2009, OishiMacLow2009}.

The radiation processes in the disk are implemented through the first (thermal diffusion as an approximation for flux limited diffusion of radiation energy density) and second (thermal relaxation to mimic heat exchange with the surface of the disk and thermal equilibration with the irradiation from the central object) terms on the right hand side of the entropy equation. As mentioned in the last chapter we keep the diffusion coefficient $K$, which is defined as in \citep{KleyBitschKlahr2009}, constant and define its value via $\tau_{\rm diff}=H^2/K$. So if the vortex has a radius of $H$, the pressure scale-hight of the disk, the diffusion time $\tau_{\rm diff}$ has the value we quote in e.g. Table \ref{setup}. If the vortex is smaller than $H$ relaxation will be much faster.

\begin{deluxetable*}{ c c c c c c  c  c c c}
\tablewidth{0pt}
\tabletypesize{\scriptsize}
\tablecaption{Simulation setup and results}
\tablehead{
\colhead{run} &
\colhead{$\beta_P$}           &
 \colhead{$\tau_{\rm cool}$ }&
  \colhead{$\tau_{\rm diff}$}&
     \colhead{$\omega^2_z$} &
   \colhead{$\alpha$}  &
  \colhead{   $u_\mathrm{RMS}$ }&
  \colhead{    $\rho_\mathrm{RMS}$}&
   \colhead{x-res}&
  \colhead{x-domain} \\
  \colhead{}&\colhead {}&
  \colhead{(${2\pi}{\Omega_0}^{-1}$)} &
  \colhead{(${2\pi}{\Omega_0}^{-1}$)} &
  \colhead{(${\Omega_0}^2$)}&
  \colhead {}&
  \colhead {($c_\mathrm{s}$)}&
  \colhead {}&
  \colhead {(gridcells$H^{-1}$)}&
    \colhead{$H$} 
  }
\startdata
A & 2.0  	  & 10  & 10& $0.056$ & $1.05\times10^{-2}$& $0.33$ & $0.22$  & 144& 4\\
A2 &  	& 10  & 10& $2.15\times10^{-2}$ & $3.09\times10^{-3}$& $0.19$ & $0.19$&144 & 8 \\
B & 		& 10  &10 & $0.060$ & $1.21\times10^{-2}$& $0.33$ &  $0.22$ &288& 4\\
C &	1.0   & 10 &10& $0.051$ & $8.67\times10^{-3}$&  $0.31$  & $0.22$& 144& 4\\
C2 &		  & 10 &10& $4.63\times10^{-3}$ & $8.2\times10^{4}$&  $0.08$  & $0.06$&144& 8\\
D &		  & 10 &10& $0.059$ & $9.63\times10^{-3}$&$0.31$  & $0.21$&288& 4\\
E &		 & 30 &10& $0.022$ & $4.33\times10^{-3}$ &  $0.23$  & $0.15$&144& 4 \\
F &		  & 30 &30& $0.022$ & $3.72\times10^{-3}$& $0.23$  & $0.15$&144& 4 \\
G &		  & 100 &10& $0.017$ & $2.61\times10^{-3}$& $0.14$  & $0.08$&144& 4 \\
H &		  & 100 &30& $0.013$ & $2.22\times10^{-3}$ & $0.15$   & $0.08$&144& 4 \\
I &	  & 100 &100& $0.010$ & $1.36\times10^{-3}$&  $0.14$  & $0.08$&144& 4 \\
J &	0.5	  &  10&10 & $5.25\times10^{-3}$ & $6.38\times10^{-4}$& $0.35$  & $0.04$&144& 4 \\
J2 &		  &  10&10 & $1.77\times10^{-3}$ & $8.91\times10^{-5}$& $0.03$  & $0.03$&144& 8 \\
K &	  &  10&10& $4.30\times10^{-3}$ & $4.30\times10^{-4}$& $0.57$ & $0.05$&288& 4\\
L &		&30 &10 &$0.021$ & $3.89\times10^{-3}$& $0.23$ & $0.15$ &144 & 4\\
M &		&  30&30& $0.021$ & $3.01\times10^{-3}$& $0.23$ & $0.15$&144& 4 \\
N &		&100&10 & $6.00\times10^{-3}$ & $1.38\times10^{-3}$& $0.14$ & $0.10$&144 & 4\\
O &		&  100&30 & $6.00\times10^{-3}$ & $1.38\times10^{-3}$& $0.14$ & $0.10$&144 & 4\\
P &		 &  100&100& $8.63\times10^{-3}$ & $1.18\times10^{-3}$& $0.15$& $0.10$&144 & 4\\
\enddata
\label{setup}
\end{deluxetable*}

To clarify that it is indeed the global entropy gradient that produces the vorticity we take the curl of the Navier-Stokes \eq{NS} and assume an equilibrium state, $u_x=0$,  and $\nabla P={\boldsymbol{0}}$ so that
\begin{equation}
\label{vort-eq}\frac{\mathcal{D}\omega_z}{\mathcal{D}t}=\frac{\beta p_0}{\rho^2R_0}\partial_y\rho.
\end{equation}
\noindent Here we see that the negative azimuthal density gradient across the vortex is the source for vorticity production proportional to the global entropy gradient.

Shearing sheet simulations with Zeus\footnote{http://www.astro.princeton.edu/\~{}jstone/zeus.html} like finite volume codes without explicit viscosity, e.g. the TRAMP code, have shown a weak amplification of kinetic energy for the pure adiabatic case, i.e. infinite cooling time (see Klahr 2013 ApJ submitted). This numerical artifact does not occur with simulations performed by the {\sc Pencil Code}. See Appendix A for a 1D radial test/comparison simulation.

Initially we apply a finite perturbation in the density so that
\begin{equation}
\rho\left(x,y\right)=\rho_0+\rho^\prime
\end{equation}
\noindent with $\rho_0$ the constant background density and $\rho^\prime$ the actual perturbation of the form
 \begin{equation}
  \rho^\prime=\rho_0Ce^{-\left(x/2\sigma\right)^2}\times\nonumber
  \sum^{k_x}_{i=-k_x}\sum_{j=0}^{k_y}\sin\left\{2\pi\left\{i\frac{x}{L_x}+j\frac{y}{L_y}+\phi_{ij}\right\}\right\},
\end{equation}
where $C$ describes the strength of the perturbation. We perturb the density in a way that $\rho_{\rm rms}=5\%$ for $\beta=1.0, 2.0$ (runs A-I) and $\rho_{\rm rms}=10\%$ for $\beta=0.5$ (runs J-P). To achieve a random perturbation we apply an arbitrary phase $\phi_{ij}$ between 0 and 1. The initial state is non-vortical. Again, this is the identical initial condition as used in Lyra \& Klahr (2011) as well as the same amplitude, $C$, for simulations with $\beta=2.0$, as was used in their simulations.

Note that with this initial perturbation we do not perturb the pressure but the entropy. Thus it is really only the term in \eq{vort-eq} that creates the development of non laminar flow structure.

All our simulations are done in dimensionless code-units. So that $R_0=\Omega_0=1$, $\gamma=1.4$, and $c_s=0.1$, which means that $H=0.1$. All time-quantities are given in $2\pi \Omega_0^{-1}$ which is one local orbit at the co-rotational radius $R_0$.

The individual setups are given in Table \ref{setup}. The thermal cooling times and thermal diffusion times are derived from  standard disk models like in \citet{Belletal1997}, also see Klahr 2013 submitted.

We explored different resolutions in our simulations, namely $288^2$, $576^2$ and $1152^2$. The unusual non power of 2 resolution comes from our computational platform with 6 core processors. Typically we used up to 24 CPUs totaling 144 cores for our largest grids. Still we needed about 1200 hours per run. The grid covers $\pm 2 H$ around $R_0$ in the radial and $[0H, 16H]$ in azimuthal direction. This leads to an effective resolution of 72 ($288^2$), 144 ($576^2$) and 288 ($1152^2$) grid-points per scale hight in radial direction and 18 ($288^2$), 36 ($576^2$) and 72 ($1152^2$) grid-points per $H$ in azimuthal direction. It is always necessary  to compromise between resolution and computational time. Lower resolution simulations are computationally less expensive but might not resolve the necessary scales. 

\section{Results}
\subsection{Saturation Values and Convergence}

\begin{figure}
\includegraphics[width=0.5\textwidth]{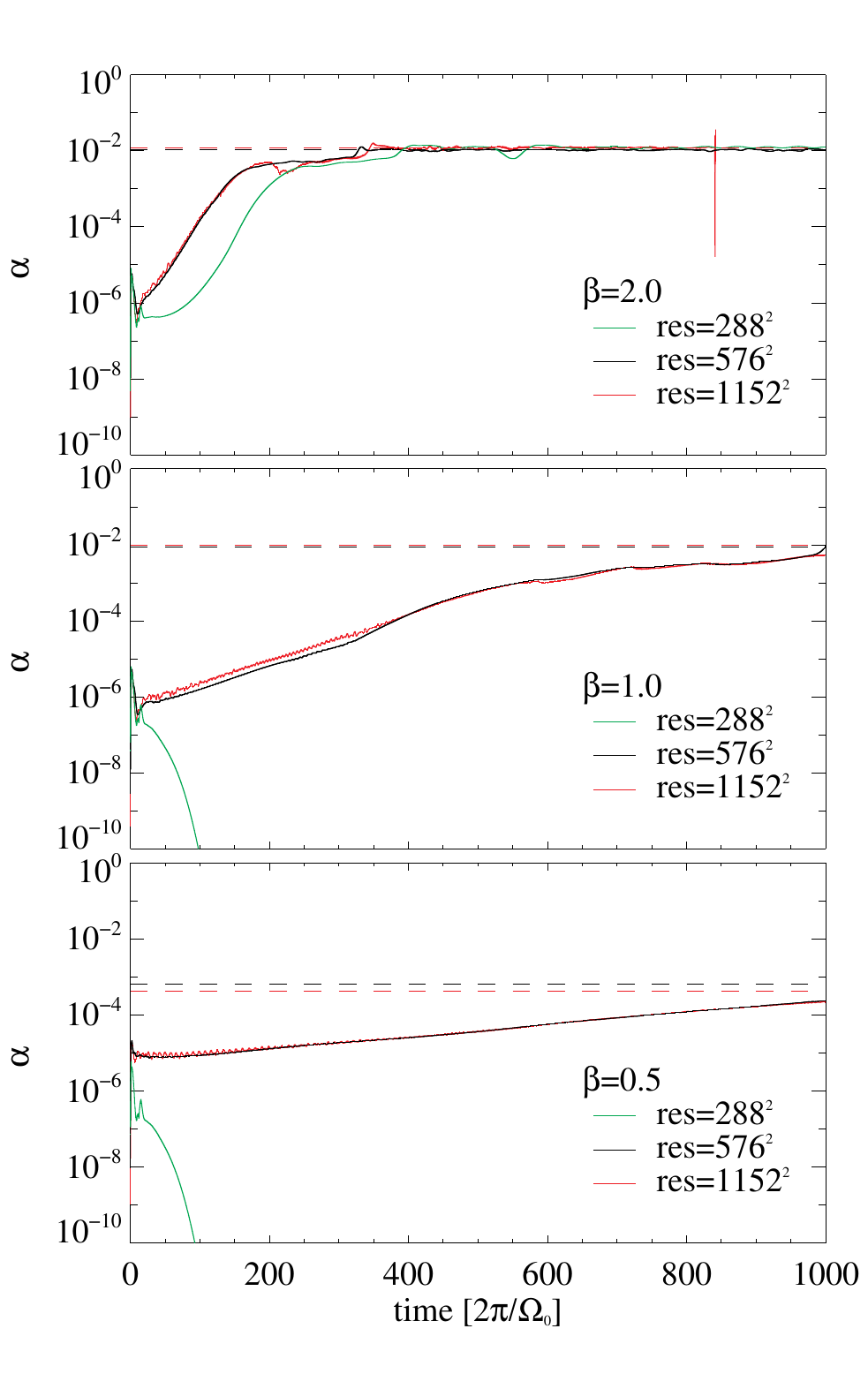}
\caption{Time evolution of $\alpha$-stresses for the three different resolutions of $288^2$, $576^2$ and $1152^2$ with an entropy gradient of $\beta=2.0$ (green line), $\beta=1.0$ (black line) and $\beta= 0.5$ (red line). For all these models $\tau_{\rm diff}=\tau_{\rm cool}=10\cdot{2\pi}/{\Omega_0}$. For all resolutions vortex amplification and therefore angular momentum transport can be seen for strong entropy gradients ($\beta=2.0$). For lower entropy gradients higher resolution is needed to see the development of vortices. The dashed lines show the saturations values ($\beta=2.0$ and $\beta=1.0$) and value at the end of the simulation ($\beta=0.5$) respectively.
}
\label{convergence}
\end{figure}

We show the time-developement of $\alpha$-stresses in \fig{convergence}. The green line shows the resolution of $288^2$, black of $576^2$ and red $1152^2$ for $\beta=2.0$ (top), $\beta=1.0$ (middle) and $\beta=0.5$ (lower panel). In all simulations $\tau_{\rm diff}=\tau_{\rm cool}=10$ local orbits.

We see that for $\beta=1.0$ and $0.5$ and a resolution of $288^2$ the perturbation decays right away. Higher resolution is required to increase the Reynolds-number of the system and have less dissipation on the smaller scales and thus excite the instability again.

We take a stronger initial perturbation for $\beta=0.5$ than for the higher $\beta$. The perturbation in entropy results in a perturbation in vorticity. This perturbation is proportional to $\beta$. For small $\beta$ we have to apply a stronger perturbation to get the same effect on the vorticity. However, we expect that if we go to even higher resolution it is possible to keep the initial density perturbation at  $\rho_{\rm rms}=5\%$ (Petersen et al. 2007).

If we compare the saturation values of runs with different resolution, we see that they differ by only 10 \% from one another (see Table \ref{setup}).

It is important to note that the instability is excited and we measure $\alpha$-values in the converged runs up to $4\times10^{-3}$ for entropy gradients as low as $\beta=0.5$. In fact, in Section \ref{dependence} we show that there is only a weak dependence of $\alpha$ on $\beta$ as $\alpha\propto\beta^{0.5}$. Fig. \ref{convergence} shows that the saturation values of $\alpha$ do not depend strongly on $\beta$, but as we will see in the next section the amplification rates do.

\subsection{Amplification- and Decay Rates}\label{secGrDe}

We analyze the amplification timescales of the vortices, meaning how fast a vortex grows due to the baroclinic feedback. Thus it is independent of the precise shape of the initial condition as long as the amplitude is large enough for the given Reynolds number to have vortex growth. In fact, the initial strong kick needed to get the vortex going decays rather quickly as can be seen in e.g. \Fig{convergence}. Here, the $\alpha$-values start out in the order of $10^{-5}$ then drop to around $10^{-8}$ as the initial perturbation decays. As soon as the baroclinic feedback sets in, the values rise again. The timespan that follows is the one where we measure the amplification time.

\begin{figure}
\includegraphics[width=0.5\textwidth]{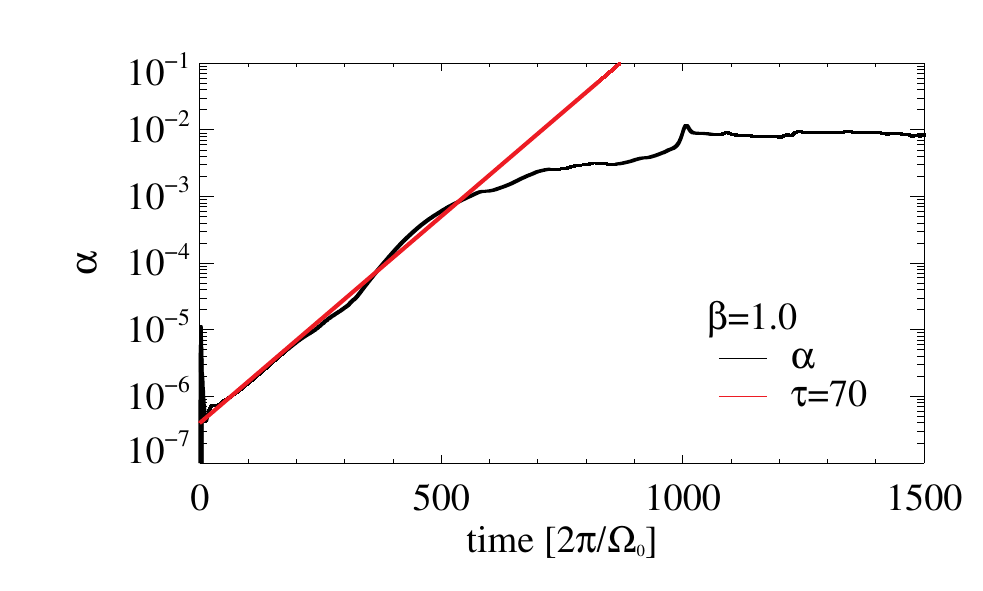}
\caption{Time evolution of the $\alpha$-values and enstrophy for $\beta=1.0$ and a resolution of $576^2$ (run C). The red slope marks exponential amplification with a amplification-time $\tau=70\frac{2\pi}{\Omega_0}$. For larger entropy gradients (smaller entropy gradients) we get faster (slower) amplification-times.}
\label{growth-fit}
\end{figure}

In analyzing the amplification-rates of the instability we find that the initial amplification-rate of the $\alpha$-stress ($\Gamma\left(\alpha\right)$), as can be seen in \fig{growth-fit} for run C, can be fitted as exponential amplification $\alpha=\alpha_0\exp\left(t/\tau\right)$ with $\tau\approx 70\beta^{-2}$. The proportionality to $\beta^{-2}$ is not what one would naively expect from a linear convective or buoyancy driven turbulence.

\begin{figure}
\includegraphics[width=0.5\textwidth]{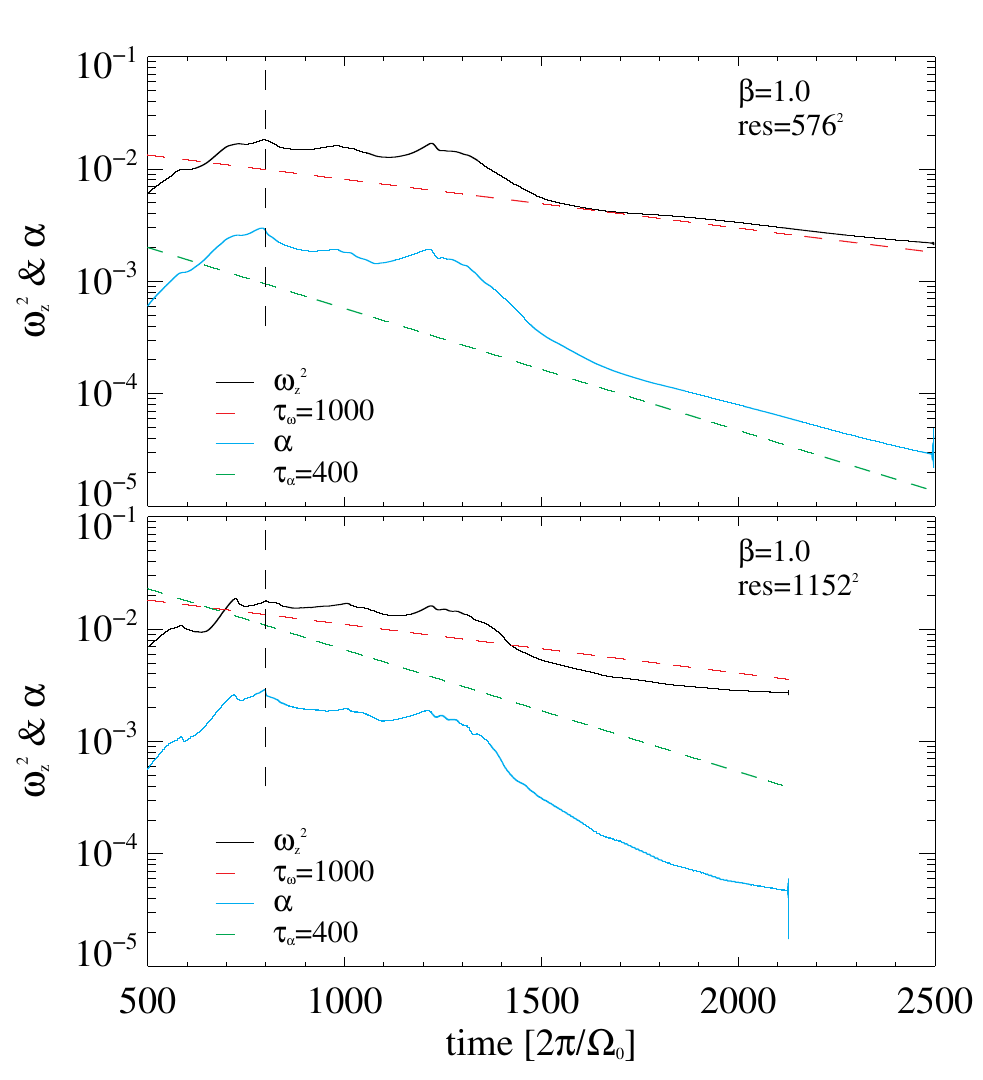}
\caption{In this run with $\beta=1.0$ a resolution of $576^2$ (run C, upper panel) and $1152^2$ (run D, lower palnel)and we turn off the entropy gradient after 800 local orbits (indicated by the black dashed line) and see how the instability decays.  Enstrophy is shown with the black line and $\alpha$-stresses with the blue line. Our fit is given through the red and green dashed lines respectively. We fit a decay time of $\tau_{\omega_z^2}=-1000$ for the enstrophy and $\tau_{\alpha}=-400$ for $\alpha.$}
\label{decay-both}
\end{figure}
For a linear buoyancy driven turbulence one would expect an amplification rate proportional to the \BV, $N$
\begin{equation}
N^2= -\frac{1}{\gamma\rho}\frac{\partial p}{\partial r}\frac{\partial}{\partial r}\ln\left(\frac{p}{\rho^\gamma}\right)
\end{equation}
\noindent which in our parameters looks like
 \begin{equation}
 N^2=-\beta_p\beta_s\frac{1}{\gamma}\left(\frac{H}{R}\right)^2\Omega^2\propto-\beta^2.
 \label{BV}
\end{equation}
\noindent Here we explicitly wrote $\beta_p$ and $\beta_s$ to make clear that the \BV \ depends on the product of entropy and pressure gradient which can be different in global simulations.

All quantities in \eq{BV} are positive. Thus the \BV \ is imaginary and therefore a linear buoyancy driven turbulence would have a amplification-rate $\Gamma\propto\ iN\propto\beta$. However, we found that $\Gamma\propto \beta^2$ provides a better fit. This once again reflects that the baroclinic vortex amplification is a non-linear effect. In linear convective instability a displaced parcel of gas feels a buoyancy force and thus accelerates propotionally to $\beta$. But in the disk baroclinic instability first a vortex has to form with an azimuthal entropy gradient proportional to $\beta$ (and $\tau_{\rm cool}$) and in a second step this vortex feels a torque proportional to $\beta$. Therefore the amplification is proportional to $\beta^2$. The $\beta^2$ and $\tau_{\rm cool}$ dependance has also been derived by \citet{LesurPapaloizou2010}, see their Eq. (23) for an order of magnitude estimate of the growthrate.

The amplification behavior in \fig{convergence} already shows convergence for 576 grid cells resolution, e.g. $144/H$ in radial direction.

If we compare our amplification timescales for the lowest entropy gradients with the migration times obtained by \citet{Paardekooperetal2010} we see that they are of the same order of magnitude. Which means that the vortex could have drifted into the central star before it reaches strong $\alpha$-values. However, \citet{Paardekooperetal2010} also state that their timescales refer to fully grown vortices of size $H$. Smaller vortices drift significantly slower. This gives them enough time to reach a size, with which they provide sufficient angular momentum transport, before they drift inward.

To study the numerical dissipation effects even further we now assess how the vortices decay if baroclinic driving is switched off (\fig{decay-both}). To do this we first evolve runs C and D with $\beta=1.0$ and the two resolutions of $576^2$ and $1152^2$ for 800 orbits and then turn off the entropy gradient so that $\beta=0.0$. We observe that the vortices get smaller and that all relevant quantities like vorticity, $\omega_z^2$, or $\alpha$-stresses decay with exponential behavior. \citet{GodonLivio1999} saw the same exponential decay of vorticity when they analyzed longevity of anti-cyclonic vortices in protoplanetary disks. Their dissipation was proportional to the effective viscosity applied in their numerical experiment. Here we find the same decay-rate for both resolutions, highlighting that the decay of vortices is no longer through numerical effects, but due to the radiation of waves as in \citet{Korotaev1997}.

\subsection{Saturation Values}

We have established that even  shallow entropy gradients lead to vortices but we still have to show that sufficient angular momentum transport can be reached with these shallow gradients. The saturation values of enstrophy, $\omega_z^2$, or $u_{\rm rms}$ are of interest as well. Note that we talk about saturation values of our 2D local simulations, where certain restrictions apply, see a more detailed discussion in the conclusions.  In the next sections we discuss the measured saturation values and analyze how the different controlling parameters influence amplification-phase and final values.

\subsubsection{Influence of Entropy Gradient}

\begin{figure}
\includegraphics[width=0.5\textwidth]{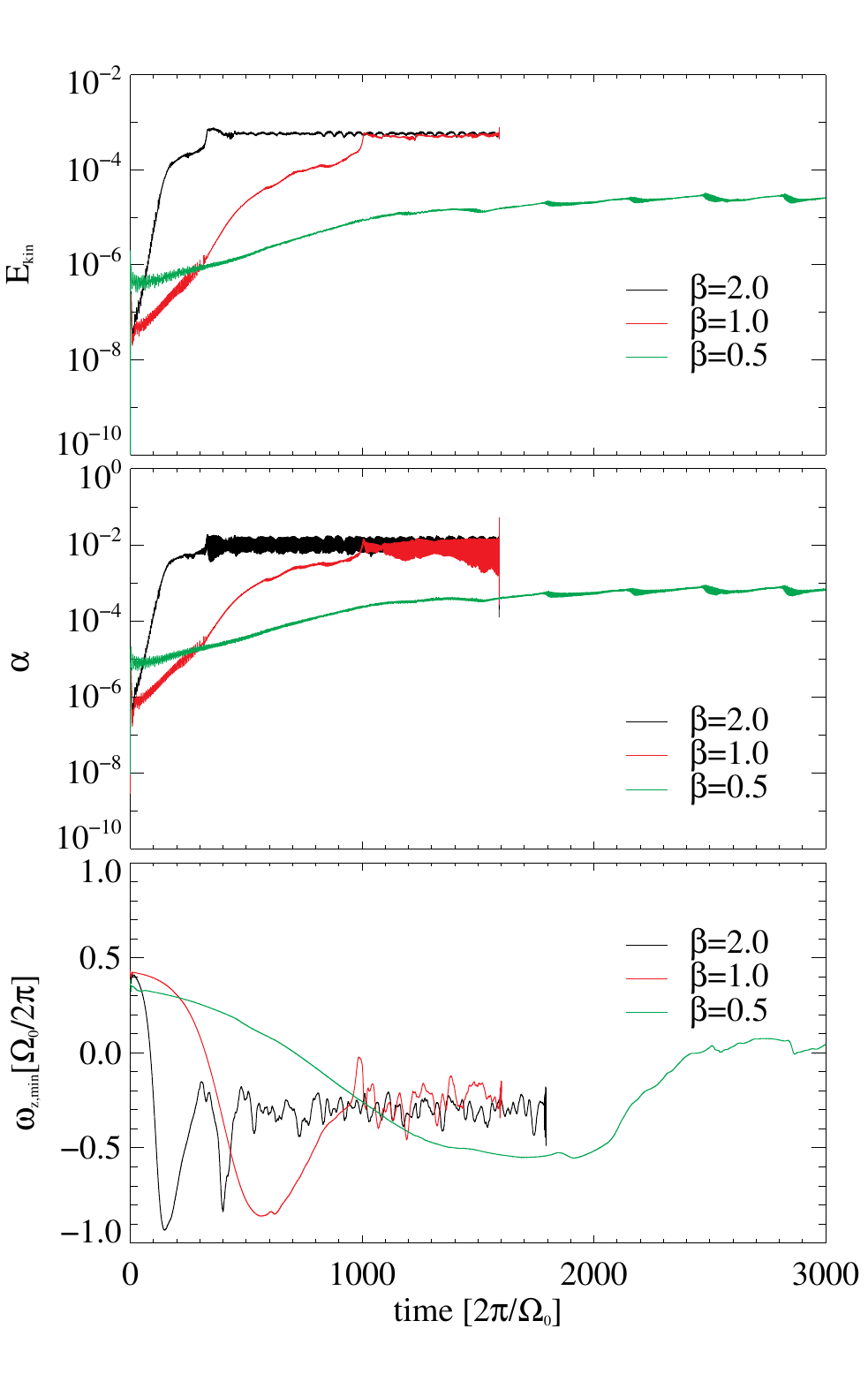}
\caption{Time evolution of kinetic energy $E_{kin}$ (top), $\alpha$-value (middle) and minimum vorticity $\omega_{z,{\rm min}}$ (bottom) for a resolution of $576^2$ and $\tau_{\rm diff}=\tau_{\rm cool}=10$ but different entropy gradientes: $\beta=2.0$ (green), $\beta=1.0$ (black) and $\beta=0.5$ (red) (runs A, C, J). Saturation is first reached for high $\beta$ already after 300 orbits, then for $\beta=1.0$ For $\beta=0.5$ no saturation is reached even after 3000 orbits. The increase in $\omega_{z,{\rm min}}$ after the point in time when saturation is reached can be explained through the heat transport across the vortex. Since it has reached its final and largest size heat transport takes longer due to the larger size of the vortex.}
\label{diff_beta}
\end{figure}

In \fig{diff_beta} we compare runs A, C and J (at a resolution of $576^2$ and $\tau_{\rm diff}=\tau_{\rm cool}=10$) which differ only regarding the value of $\beta$.  There is an initial exponential amplification-phase of $\alpha, E_{\rm kin}$ and $\omega_z^2$ that is shorter for high $\beta$, followed by a saturated state. We also see that for lower $\beta$ the saturation values are lower. We want to stress that we did not reach saturation for simulations J and K (at a resolution of $576^2$ and $1152^2$ and $\tau_{\rm diff}=\tau_{\rm cool}=10$). Even after 3000 local orbits vortex amplification was still ongoing. Here, $\tau_{\rm diff}=10$ is much shorter than the amplification-rate we estimated in the previous section ($\tau\approx300$). As we will see in the next section the amplification-phase is shortest if those time-scales are comparable, because $\tau_{\rm diff}$ also defines how fast pressure perturbations are damped. Although we expect the saturation values of simulation J and K to be higher than what they are right now, it is possible that they will still stay below the saturation values obtained in simulations with higher $\beta$.

The vorticity can be seen as a measure of the strength of the vortex. The higher the absolute value of the vorticity the stronger the vortex. The only stable vortices in disks are anticyclonic\footnote{Cyclonic vortices are also possible, but are quickly destroyed by shear \citep{GodonLivio1999}.} and therefore the vorticity has negative values. So the minimum value of vorticity ($\omega_{z, \rm min}$) shows how strong a vortex is. To explain the behavior of $\omega_{z, \rm min}$ (3rd panel in \fig{diff_beta}), cooling processes have to be taken into account. During the early phases thermalization is dominated by thermal diffusion \citep{PetersenStewartJulien2007}. As mentioned before this time-scale is shorter for smaller vortices. Therefore heat exchange between the vortex gas and the ambient gas is more efficient than in later stages. Once the vortex has grown to its final size, thermal relaxation takes over. However heat exchange in the center of the vortex is less efficient than in the earlier stages. The baroclinic feedback, e.g. the azimuthal entropy gradient across the vortex,  is less efficient, the vortex grows weaker, and $\omega_{z, \rm min}$ rises again, creating a flat yet extended vortex.

\subsubsection{Influence of Thermal Diffusion and Cooling Times}

\begin{figure}
\includegraphics[width=0.5\textwidth]{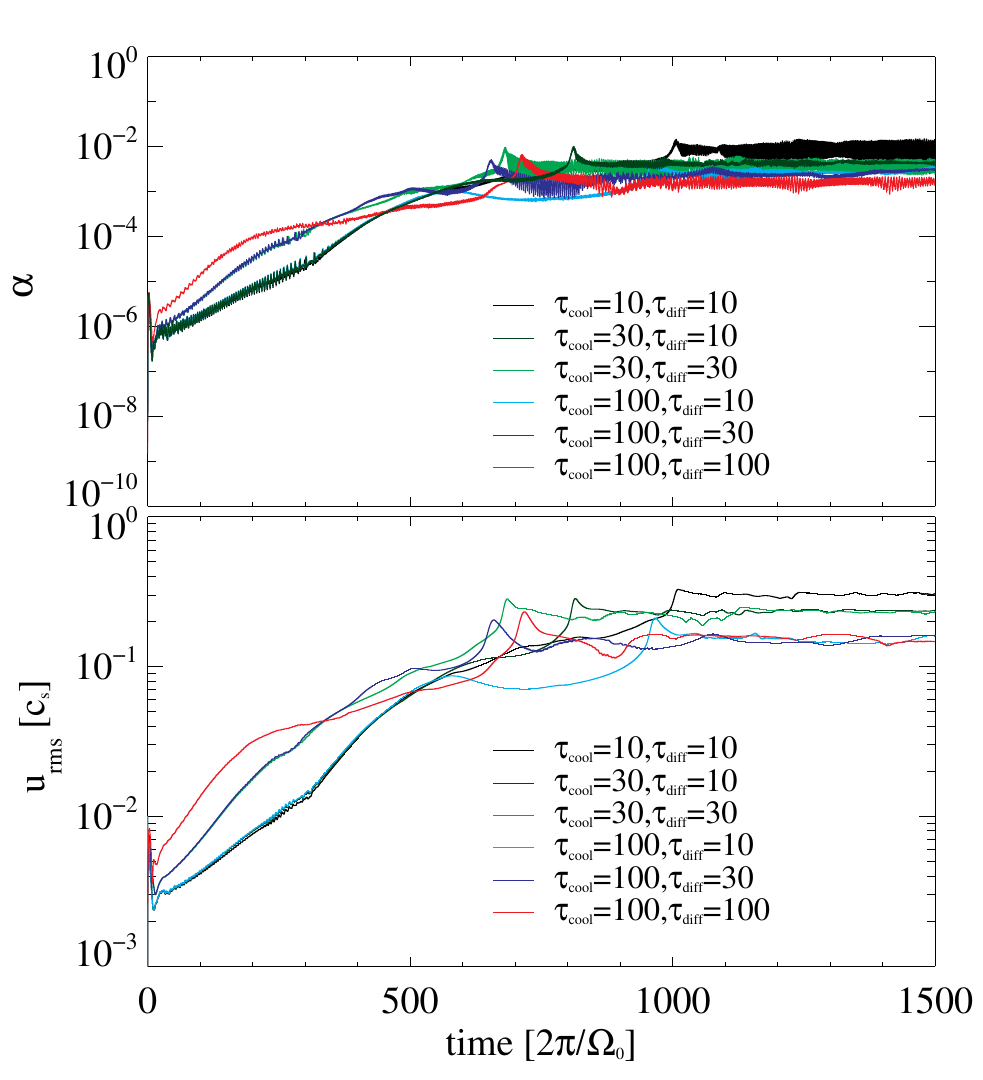}
\caption{Comparison of different $\tau_{\rm diff}$ (right numbers) and $\tau_{\rm cool}$ (left numbers) for same $\beta=1.0$ (Runs C-I). The top panel shows the $\alpha$-value and the bottom one $u_{\rm rms}$. One can see that the early amplification-phase is determined by the diffusion time since the heating across the vortex is more important then vertical heat transport. We get faster amplification for higher $\tau_{\rm diff}$. Once the vortex grows larger heat transport gets more difficult and thermal relaxation dominates. Therefore the saturation values are determined through $\tau_{\rm cool}$. Saturation values are higher for shorter $\tau_{\rm cool.}$}
\label{diff_tau}
\end{figure}

We take a closer look at simulations with $\beta=1$ and different combinations of $K$ and $\tau_{\rm cool}$ to see how thermal diffusion and relaxation influence the saturation values and the amplification-phases. As long as $\tau_{{\rm diff} (l)}=l^2/K < \tau_{\rm cool}$, $\tau_{{\rm diff} (l)}$ will dominate the heat exchange from the inside of the vortex to the ambient disk. As the vortex grows $\tau_{{\rm diff} (l)}$ will increase and with that only contribute to the heat exchange at the outskirts of the vortex. $\tau_{\rm cool}$ will then dominate the interior of the vortex.

For the simulations where we set $\tau_{\rm diff}= \tau_{\rm cool}$, $\tau_{\rm cool}$ will take over when the vortex has reached a size of $H$. In radial extend this happens once the vortex has grown to its final size. 

This is consistent with what we see in \fig{diff_tau}. During the early amplification-phase simulations with equal $\tau_{\rm diff}$ behave exactly the same. Eventually $\tau_{\rm cool}$ takes over so that the saturation values are determined by $\tau_{\rm cool}$. For longer $\tau_{\rm cool}$ saturation values are lower than for shorter $\tau_{\rm cool}$.

\subsubsection{Influence of Physical Domain}

\begin{figure}
\includegraphics[width=0.23\textwidth]{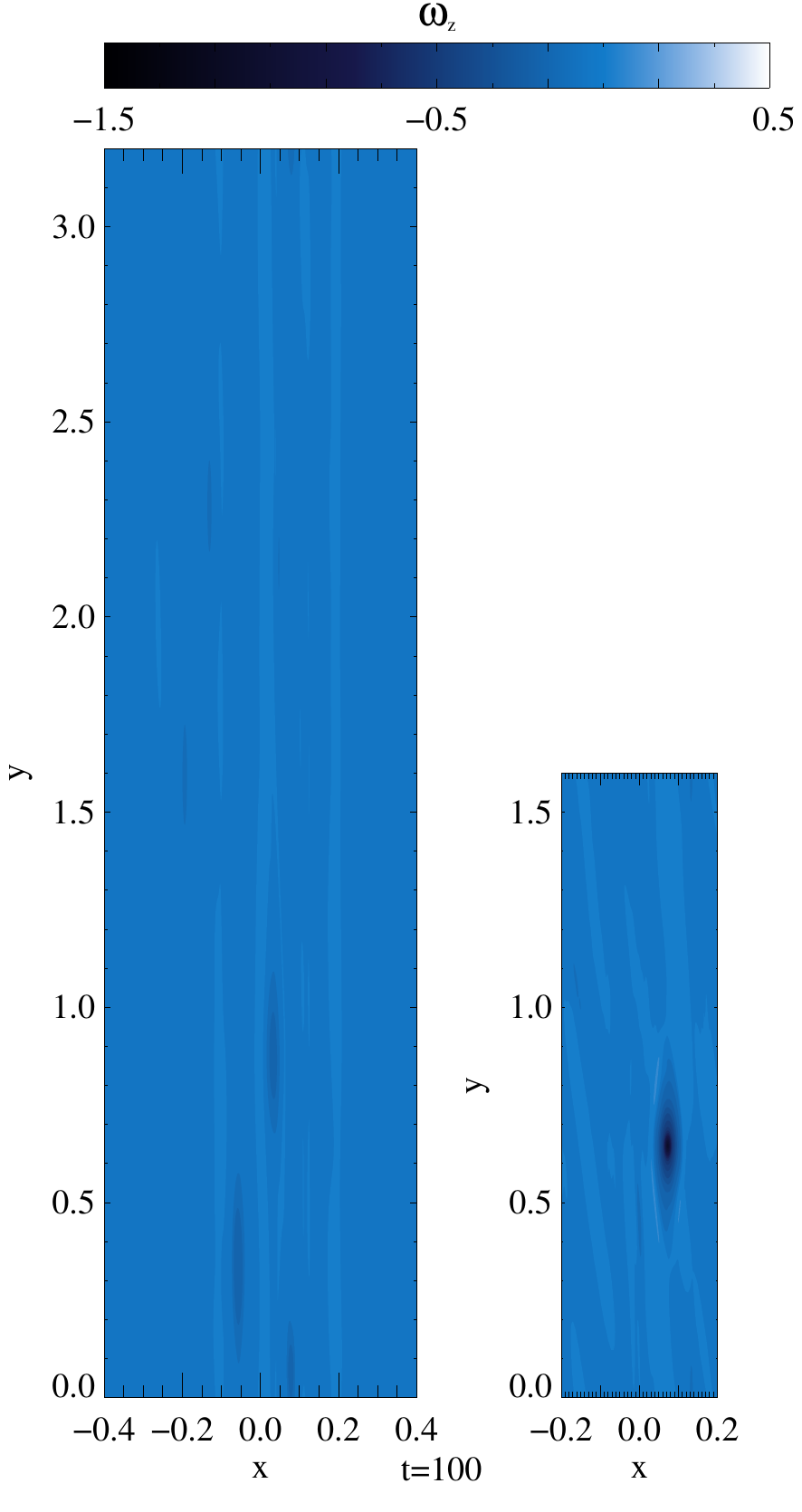}
\includegraphics[width=0.23\textwidth]{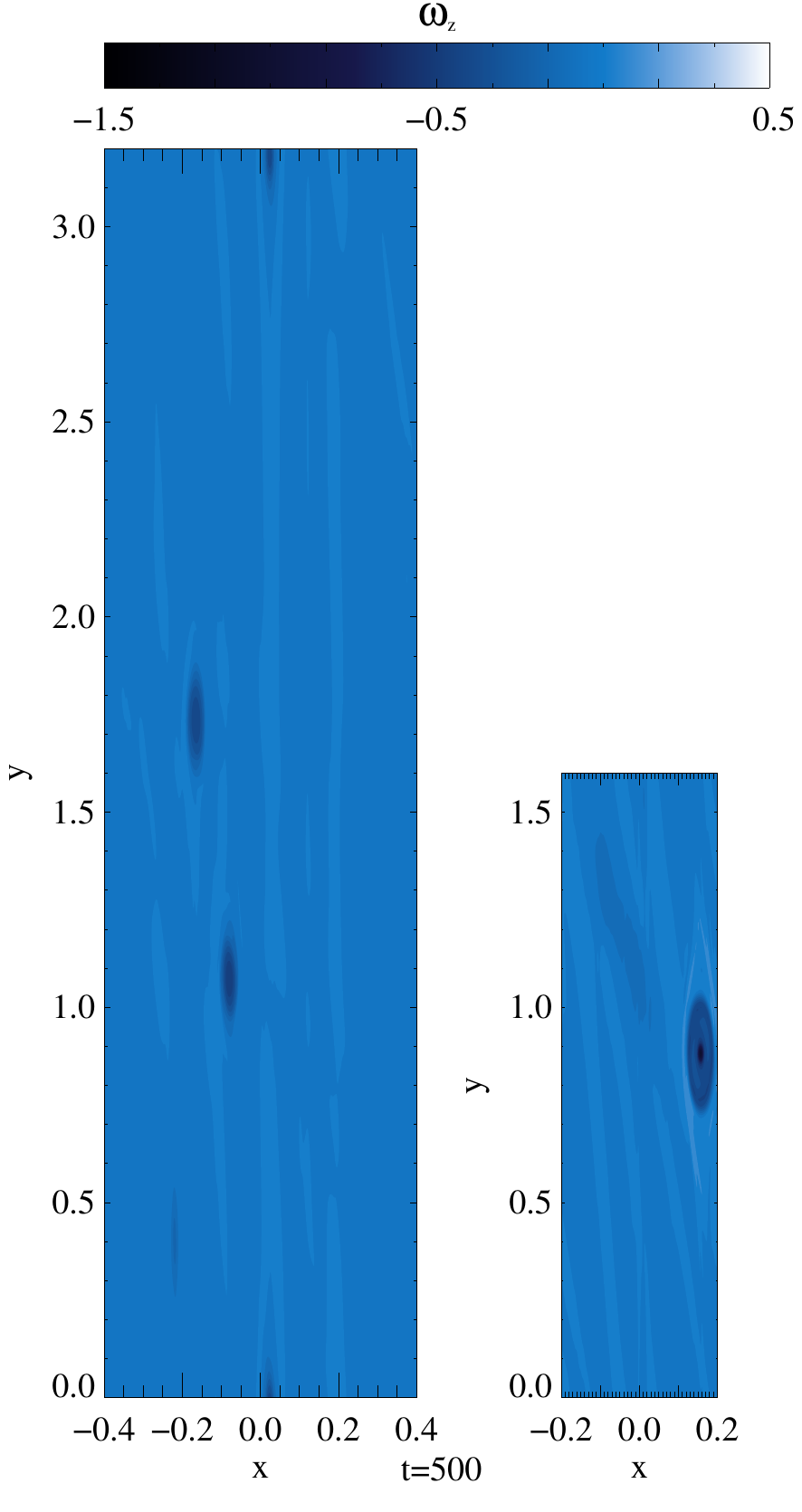}\\
\includegraphics[width=0.23\textwidth]{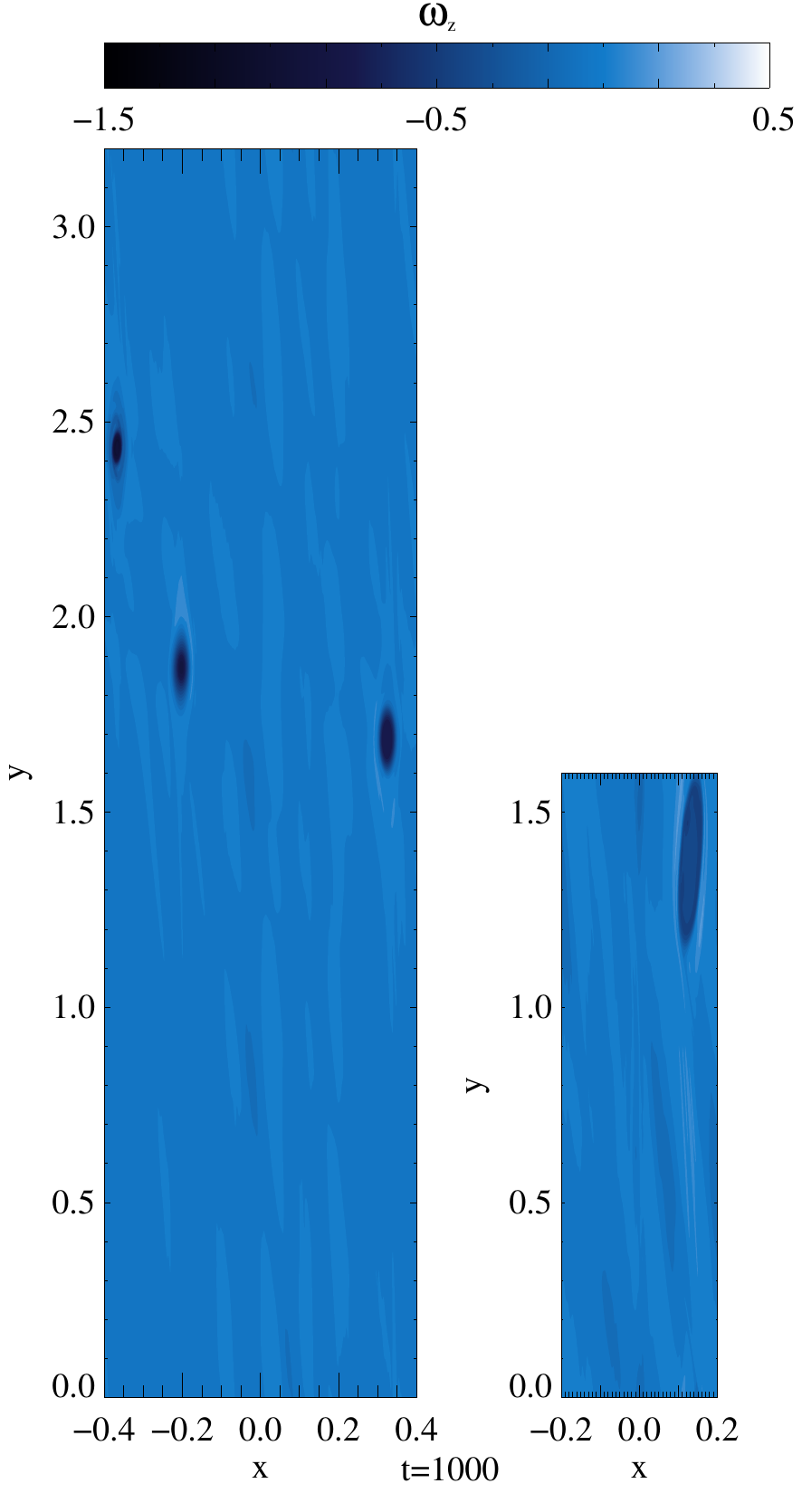}
\includegraphics[width=0.23\textwidth]{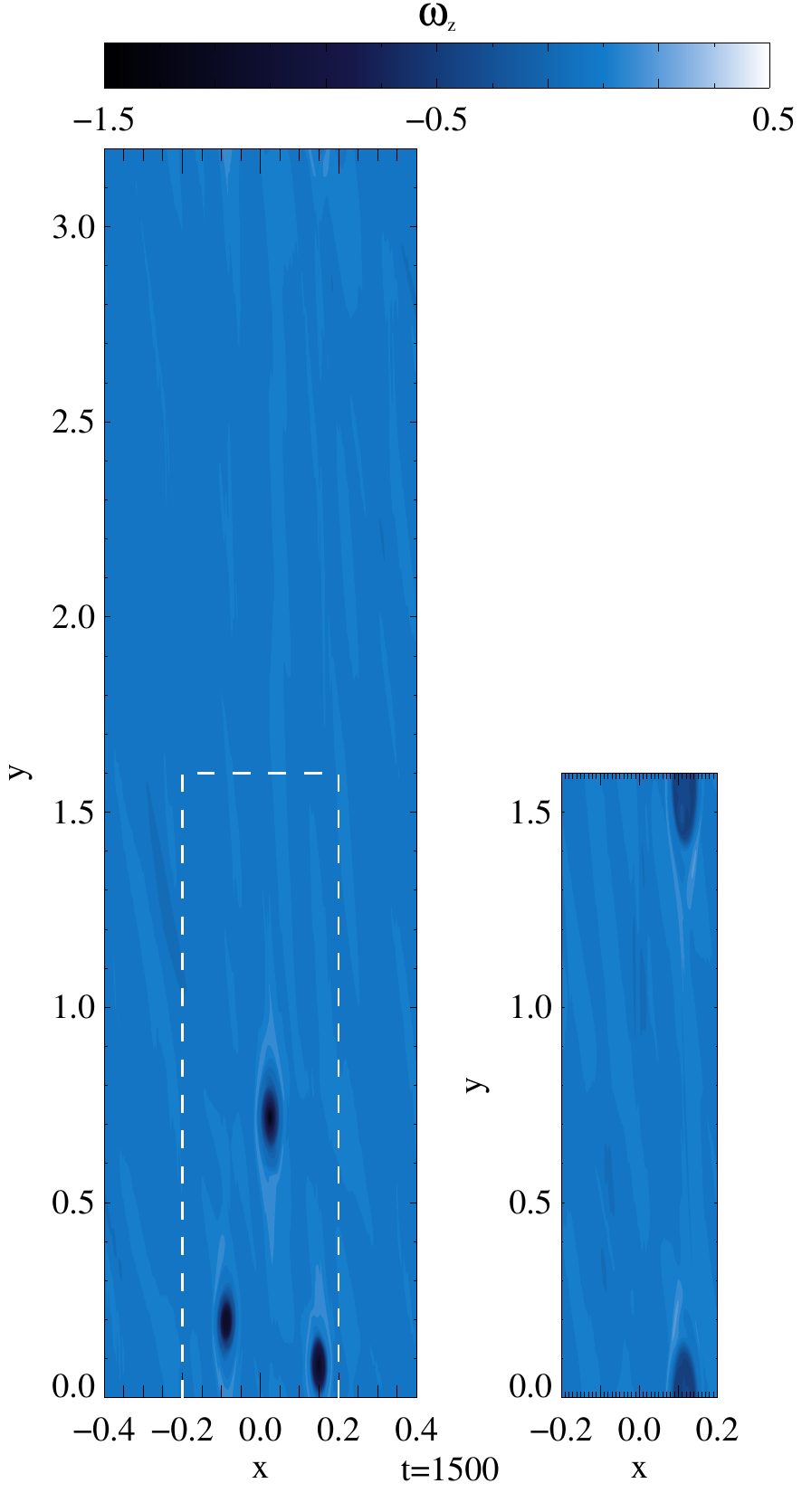}
\caption{Snapshots of the z-component of the vorticity, $\omega_z$ after 100, 500, 1000, 1500 local orbits for the two different physical domains with $\beta=0.5$. Initially both runs have vortices of equal size. Since there is less space between vortices, they can merge sooner in runs with the small physical domain. The vortices in the large physical domain take longer to grow. The dashed white box in the last plot indicates the area of the small physical domain.}
\label{planes05}
\end{figure}

\begin{figure}
\includegraphics[width=0.5\textwidth]{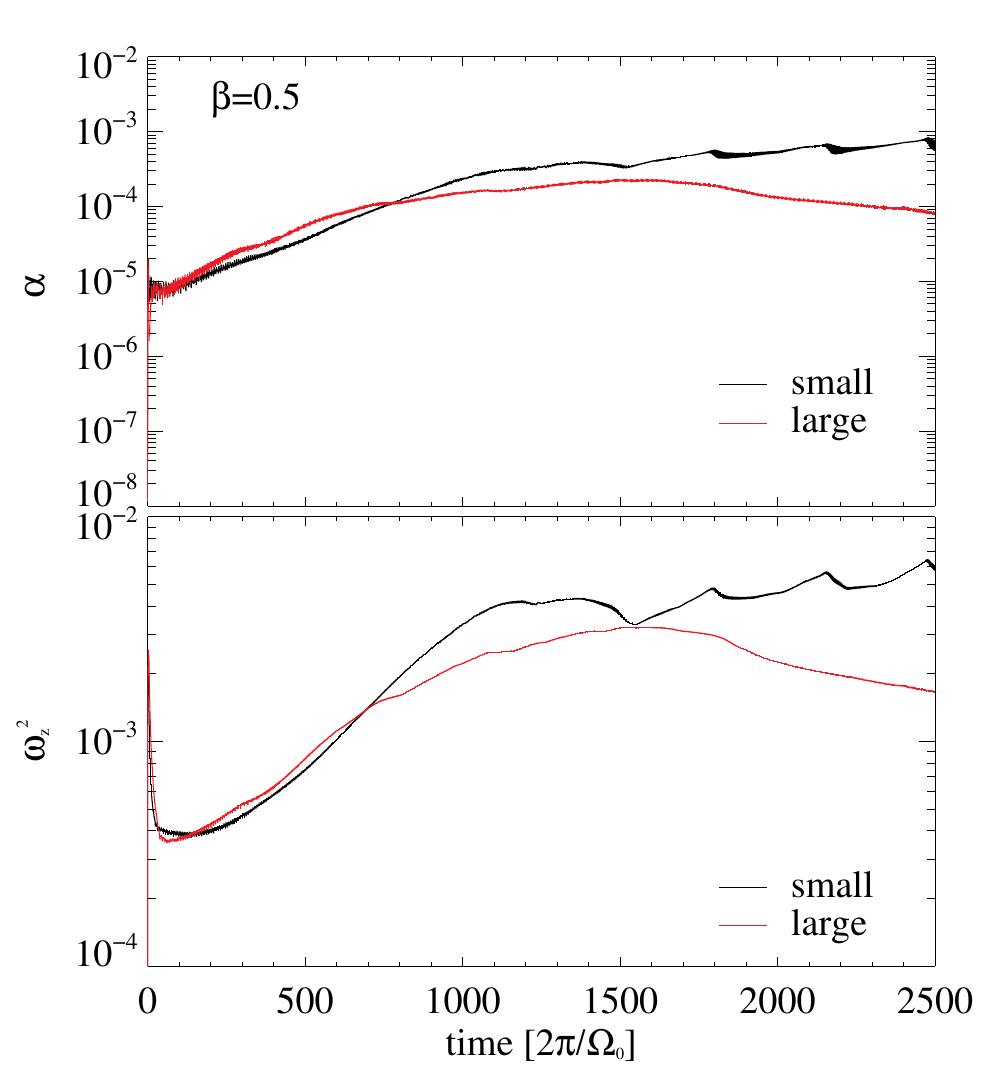}
\caption{Time development of $\alpha$ and $\omega_z^2$ with $\beta=0.5$ for small (black) and large (red) physical domain (runs J and J2). Saturation values are lower in the large box than in the smaller box.}
\label{large_b05}
\end{figure}

\begin{figure*}
\includegraphics[width=\textwidth]{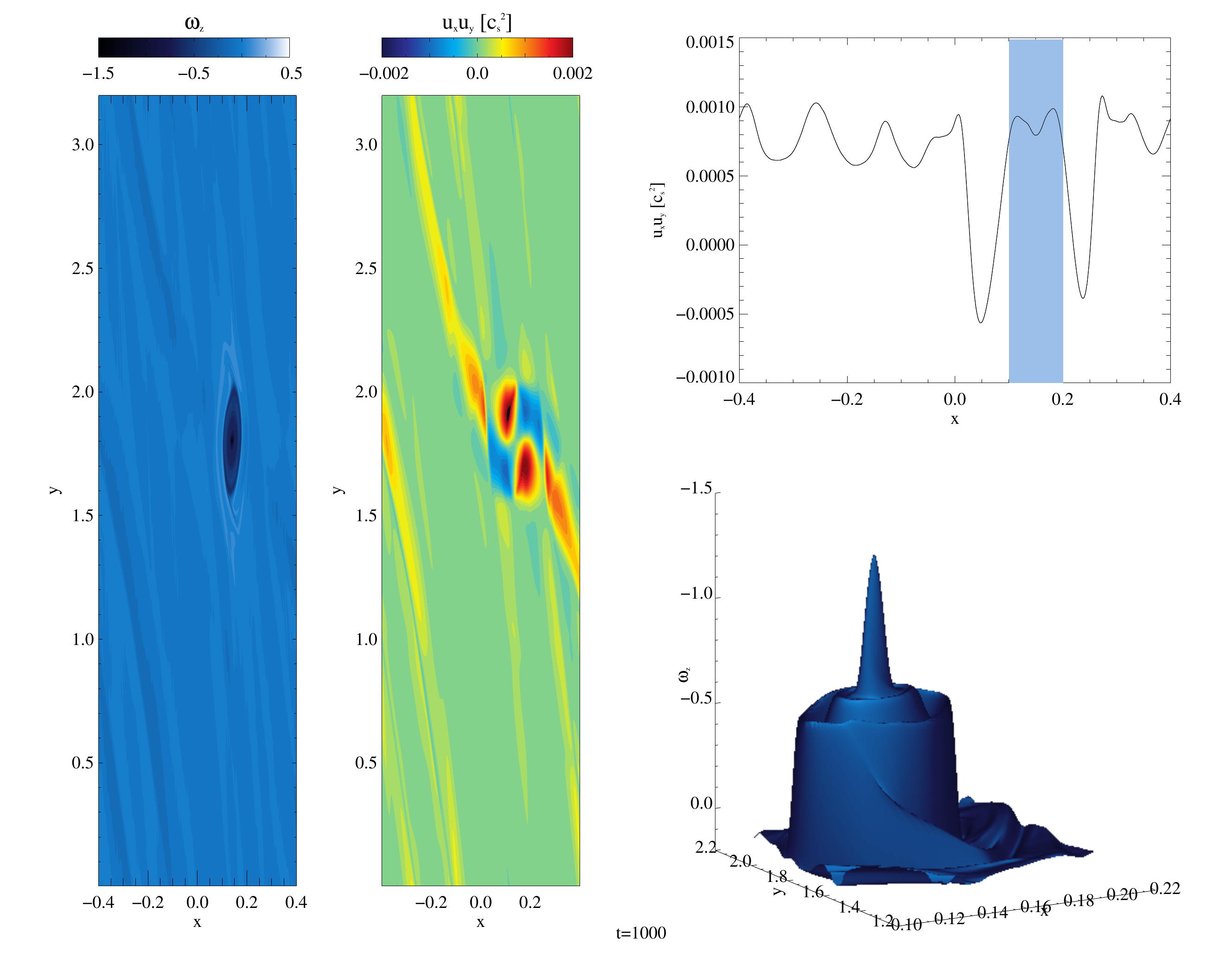}
\caption{Vorticity profile (left) and $\alpha$-stress (middle) for $\beta=1.0$ and the large physical domain (run C2). Yellow and red areas denote positive $\alpha$-vaues whereas blue areas show negative $\alpha$-stresses. In green areas $\alpha=0$. One can see the waves excited by the vortex. Those waves are responsible for the angular momentum transport. It is a localized process. Since the vortex and the vorticity-waves fill out a smaller area of the box in the large box (large green areas where there is no angular momentum transport) and our calculation of the saturation values averages over the entire area of the box, the saturation values seem to be lower. The plot in the right upper panel shows an azimuthal average over the $u_xu_y$. Inside an ideal vortex $\alpha$-stresses would sum up to zero. However, as indicated in the lower right plot, the vortex has a complex structure which leads to deviations from the idealized shape.}
\label{large_alpha}
\end{figure*}

A problem with local shearing sheet simulations is that eventually vortices grow to box-size. We cannot say whether they have reached their final size or just do not have any more room to grow. Another problem that arises with the periodic boundary conditions is that the vortices potentially interact with themselves and thus forcing (shaking) them to shed more waves and therefore increase the $\alpha$-values. To deal with that, we re-did simulations A, C and J with a doubled physical domain (simulations A2, C2, J2 in Table \ref{setup}). The resolution is the same. Instead of $x=\left [-0.2, 0.2\right ]$ and $y=\left[0.0, 1.6\right]$ we switch to $x=\left [-0.4, 0.4\right ]$ and $y=\left[0.0, 3.2\right]$. We did not adjust the initial perturbation in any way. Therefore the initial state is perturbed at smaller wave numbers than in the smaller domain. If we go to even larger boxes the initial condition has to be adjusted so the the effective perturbation in the density is of the same strength as in the smaller physical domain.

If we compare the time development of runs with a different physical domain (see \fig{planes05}), we see that vortices in fact do not merge as fast in the large domain because there now is more space between them in radial direction, and they thus pass each other less frequently due to the extended azimuthal domain. Eventually they can merge as \citet{GodonLivio1999} saw, but the larger the box the longer it takes.  We do not want to discuss the mechanism of how the process of vortex merging happens exactly. This has been explained extensively in the field of fluid dynamics \citep[see e.g.][]{CerretelliWilliamson2003}. The merging process itself is not the focus of our study, because a) the vortex merging is strongly influenced by the box dimensions in a shearing sheet simulation and b) 2D flat vortices merge differently than full scale 3D vortices. The important thing is that vortices do indeed merge if the are sufficiently close to one another, but conserve $\omega$ in the process.

Another unphysical process that can occur in local periodic simulations is that  when the vortex approaches the integral scale it interacts with itself, the outer edges of the one side of the vortex almost touches the other side of the same vortex. We do not see this for the runs with the larger physical domain. Since the vortices in the larger domain do not interact with themselves, the saturation values are lower. However, they are still in the same order of magnitude (see Table \ref{setup}).

In \fig{planes05} we show snapshots of the vorticity for $\beta=0.5$ (simulations J and J2). Initially there are several vortices. The larger ones sweep up the smaller vortices and thus grow further. At 1500 local orbits there is only one vortex left for the small physical domain, whereas in the larger physical domain there are still three vortices.

If we look at the $\alpha$-value and enstrophy for these two simulations (see \fig{large_b05}) we see that the value seems to decay in the larger box at the end of the run. However this does not mean that the vortices die out. It more so reflects fluctuations in the vortex interaction, modulating $\alpha$, as also can be seen in the small domain case at high frequency. We calculate the values as a mean over the entire box but especially the angular momentum transport is a very localized process as can be seen in \fig{large_alpha} (this time for $\beta=1.0$ after 1000 orbits). Here we show the product $u_xu_y$ at each location in the box. Most areas of the box have an $u_xu_y$-value close to zero. However, one can clearly see bands excited by the vortex with positive $u_xu_y$-values. These bands are inertia-acoustic waves which are responsible for the angular momentum transport \citep{KlahrBodenheimer2003, MamatsashviliChagelishvili2007, HeinemannPapaloizou2009, Tevzadzeetal2010}. If we had an ideal vortex with a smooth surface we would expect that $u_xu_y$ sums up to zero within the vortex. However the vortex has a more complex structure as can be seen in the lower right plot of \fig{large_alpha}. This leads to an negative net $\alpha$-value across the vortex.

To properly compare the values of $\alpha$ for both physical domains, the box average has to be taken. If the average over an equal physical size centered around a vortex, as indicated by the white dashed lines in Fig. \ref{planes05}, is taken, then the $\alpha$-values agree again. The $\alpha$-values are generated only in the vicinity of vortices.

\subsection{Correlations}

\begin{figure}
\includegraphics[width=0.5\textwidth]{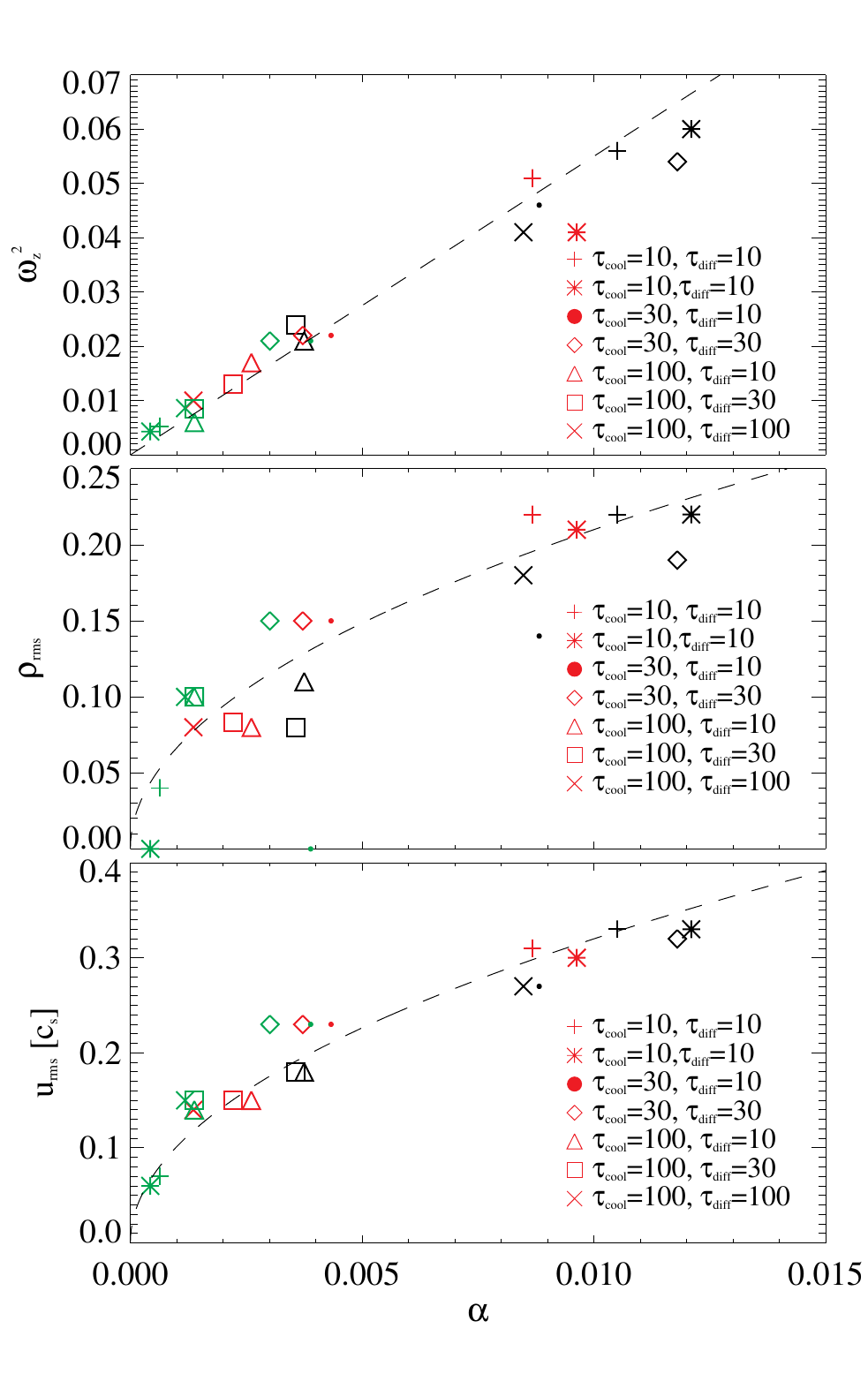}
\caption{Saturation values of $\omega_z^2$, $\rho_{\rm rms}$ and $u_{\rm rms}$ as a function of saturated (value at the end of the simulation for $\beta=0.5$) $\alpha$-value and all our runs with the small physical domain (runs A-P). The symbols show the different combinations of $\tau_{\rm cool}$ (left numbers) and $\tau_{\rm diff}$ (right numbers). Where red are runs with black $\beta=2.0$, $\beta=1.0$ and green $\beta=0.5$. The black dashed line shows the dependency that we fit.}
\label{a-dep}
\end{figure}

It is a feature of baroclinic instability that the saturation values of $u_{\rm rms}$, $\omega_z^2$, $\rho_{\rm rms}$ seem to correlate with each other. In \fig{a-dep} we plot the different quantities as a function of $\alpha$. Figure \ref{a-dep} shows the dependencies on $\alpha$ for all our simulations. The colors represent the different entropy gradients: $\beta=2.0$ (black), $\beta=1.0$ (red) and $\beta=0.5$ (green). The different combinations of diffusion and cooling times are represented through the different symbols. We find that the following relations are good fits to our simulation results
\begin{eqnarray}
{u}_{\rm rms}&=&3\sqrt{\alpha}c_s\label{alpha_vrms} \\
\rho_{\rm rms}&=&2\sqrt{\alpha}\rho_0\\
\omega_z^2&=&5\alpha\Omega_0^2.
\end{eqnarray}
We can derive the typical length-scale of angular momentum transport $L$, of the system if \eq{alpha_vrms} is inserted into the general $\alpha$ formalisms \citep{ShakuraSunyaev1973} $\nu=\alpha c_sH=u_{\rm rms}L$
so that
\begin{equation}
L=\frac{\sqrt{\alpha}H}{3},
\end{equation}
\noindent indicating smaller structures than the vortices in our simulations and also smaller than the vorticity in standard $\alpha$-models where $\omega\propto\sqrt{\alpha}$ with a different coefficient \citep{Cuzzietal1994}.

We do not perform a more exact analysis of these dependencies (varying initial conditions) before we do three-dimensional simulations.

\subsection{Dependence on $\beta$}\label{dependence}

\begin{figure}
\includegraphics[width=0.5\textwidth]{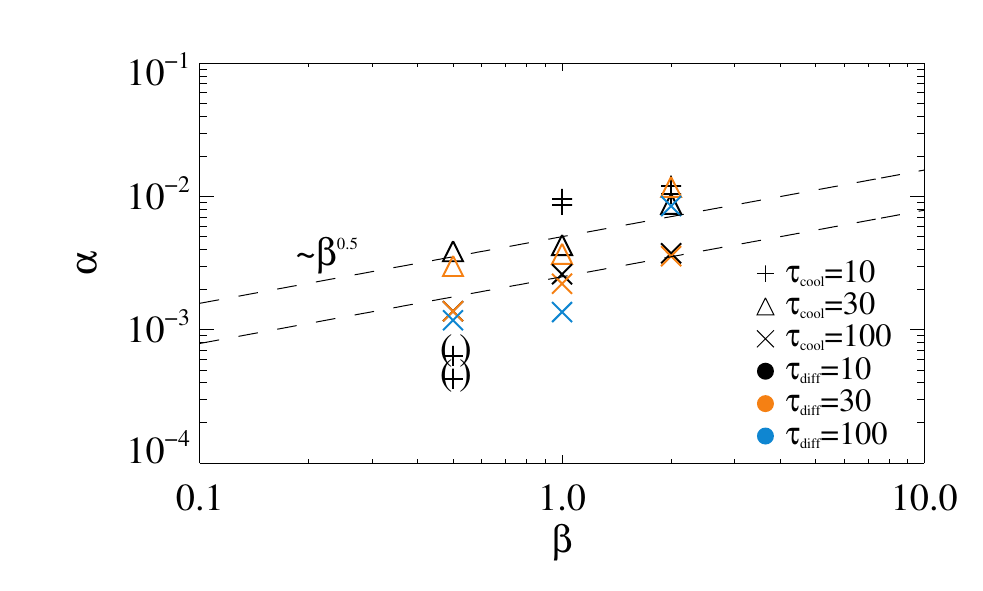}
\caption{Saturation values of $\alpha$ for all our runs with the smaller box depending on $\beta$. Runs with parentheses around them were not saturated at the end of the simulations. Therefore we do not take them into account when we fit the $\alpha-\beta$-relation.The symbols show the different combinations of $\tau_{\rm cool}$ (symbols) and $\tau_{\rm diff}$ (colors). }
\label{a_beta}
\end{figure}

In Section \ref{secGrDe} we showed that amplification of vortices for low entropy gradients is computationally demanding in terms of evolution time. Thus it is difficult to extract saturation values for entropy gradients even shallower than $\beta=0.5$ with the computational resources at hand.

In \fig{a_beta} we plot the $\alpha$-stresses as a function of the entropy gradient. Note that we choose a different color-coding than in \fig{a-dep}. Here symbols represent the thermal cooling times whereas colors represent thermal diffusion times. The dashed black line illustrates a slope $\propto\beta^{0.5}$ which is a reasonable fit for the set of points with $\tau_{\rm cool}=30, \tau_{\rm diff}=10$ (black triangles) and $\tau_{\rm cool}=100, \tau_{\rm diff}=30$ (orange x). We cannot predict $\alpha$-values for specific entropy gradients and thermal cooling and relaxation times.

The key issue is less a strong correlation between $\alpha$ and $\beta$ but rather the lack thereof. The strength of the $\alpha$-stresses reflects the size and the amplitude of the largest vortex. Its size is defined by $H$ only and not by any of the other $\tau$ and $\beta$ parameters. As long as $\tau$ and $\beta$ are sufficient to replenish vorticity at the loss-rate, the $\alpha$-stresses should be independent of $\tau$ and $\beta$. The loss time-scale via generation of waves and Reynolds stresses is rather long, see Section \ref{secGrDe} and \fig{decay-both}. Thus as long as the amplification-rates are faster than decay-rates one should always obtain roughly the same $\alpha$-values.

\section{Summary and Conclusion}

In this paper we have conducted an extensive parameter analysis for the baroclinic vortex amplification. In particular we analyzed the influence of the global entropy gradient, thermal relaxation and cooling as well as numerical parameters such as resolution, box size, and amplification-rates for vortices and saturation values of $\alpha$.

The most important result of our study is that we find vortex growth even for entropy gradients as low as $\beta=0.5$. However the amplification rate is of the order of several 100 local orbits which makes it difficult to extract reliable saturation values for the efficiency of angular momentum transport. 

Recently \citet{Paardekooperetal2010} studied the migration behavior of vortices in global accretion disks. They found significant radial drift for fully grown vortices with drift times shorter than the vortex amplification times we measure in this paper. Nevertheless, this is not a contradiction, because as also shown in \citet{Paardekooperetal2010} drift rates strongly depend on vortex size. Thus the typical life cycle of a growing vortex might be starting as a growing small vortex without relevant radial drift, which starts drifting as soon as it reaches its saturated state. Therefore radial drift does not affect the study of vortex amplification discussed here. However, it will affect the time a single vortex can partake in angular momentum transport. Future work will have to investigate radial drift of growing vortices in global simulations. Note here that \citet{Paardekooperetal2010} studied the migration in barotropic disks, in which no vortex amplification occurs.

The amplification-phase of the vorticies can be measured in the strength of the overall velocity fluctuation which seem to be growing exponentially on a certain time-scale $\tau\propto\beta^{-2}$. Therefore amplification for steeper entropy gradients is faster, i.e. $\tau=16$ for $\beta=2.0$ and $\tau=70$ for $\beta=1.0$. With these short amplification-times we do reach saturation. Whereas the $\beta=0.5$ was still growing after 3000 orbital periods, when we stopped the simulation.

Other parameters that influence the evolution of $\alpha$-stresses are the thermal cooling and relaxation times. The diffusion times define the amplification phase of the vortices because diffusion dominates small scales, e.g. small vortices.  We see faster amplification for longer diffusion times. Cooling time on the other hand determines the saturation values. Here, longer time-scales produce lower saturation values.

For the angular momentum transport we get $\alpha$-values up to $10^{-2}$ for $\beta=2.0$ and $10^{-3}$ for $\beta=1.0$ and $\beta=0.5$. These values are not so different to the ones found with MRI in active layers \citep{Flocketal2011} and stronger than the $10^{-4}$ found in dead zones \citep{Nataliaetal2010}, which shows that entropy gradients can be an important mechanism to transport angular momentum in a dead-zone. Realistic entropy gradients in protoplanetary disks are around $\beta=0.5$ and $\beta=1.0$ which can be derived out of the data obtained by \citet{Andrewsetal2009} as discussed in Klahr (2013 submitted to ApJ). Although we could not reach saturation in all our simulations for these entropy gradients we do see reasonable $\alpha$-stresses of the order of $10^{-3}$ to $10^{-2}$. We expect the final values to be in this range which still provides sufficient angular momentum transport in a disk. Yet, we have to consider certain cavities: 1.) Our simulations are 2D simulations and lack the 3 dimensional structure of the vortices. This might very well affect the strength of the $\alpha$-values. 2.) We do not consider migration of vortices, but rather have periodic boundary conditions. It is not clear for how long vortices can play a role in angular momentum transport before they migrate into the central star. Thus we cannot say how many vortices are in a disk at any given time. The higher the number of vortices, the higher the $\alpha$-values will be. The interplay between migration and Reynolds stresses definitely has to be analyzed in future models. 3.) The formation process for vortices is still not clear. It is unknown how long the initial formation of a vortex takes, by which process they are formed and if there are processes which can destroy them before the reach full growth. Therefore, our saturation values have to be viewed with caution and cannot be seen as face values for protoplanetary accretion disks. As relation between entropy gradient and strength of angular momentum transport we only find a weak dependence of $\alpha \propto \beta^{1/2}$.

Since local simulations are always limited by the box size we also conduct simulations in larger boxes. We do not see a difference in the initial amplification-phase. At later stages the amplification last longer for larger boxes and also is slower. Since part of the vortex evolution happens through merging of smaller vortices, growth takes longer in larger boxes simply because there the radial distance between vortices is bigger and thus mergers are less likely.

The saturation values of velocity fluctuations reached for the larger box sizes are slightly lower than for the smaller box sizes. This is due to two reasons. One is that we see some artificial enhancement in vortex strength in the smaller box. Once the vortex has reached box-size it can no longer grow. It is forced to interact with itself thus emitting more waves. This does not happen in larger boxes.

The other reason is that the number of vortices per radial distance is independent of box size because their typical maximum size is in the order of a pressure scale-height. In the azimuthal direction the number of vortices is limited to 1 per radius, because otherwise merging will occur on short time-scales. Therefore the overall density of vortices per simulation volume (area) is lower in simulations with the larger azimuthal extend. Here we want to note that our larger boxes with $H/r = 0.1$ and $L_y=32$ are only a factor of about two shy of the equivalent $2\pi$ global simulation.

Overall, we conclude that the baroclinic vortex amplification works reasonably well for entropy gradients as low as $\beta=0.5$. This $\beta$ corresponds to a Richardson-number of $Ri=-1.5\times10^{-3}$. This makes BVA a relevant mechanism for angular momentum transport in the dead-zone.

An exploration of lower entropy values will have to be postponed due to the long evolution time required. In the future we will study stratified 3D boxes and the interaction of dust with the vortices.

\acknowledgements

{Our simulations were conducted partly on the MPIA cluster THEO in Garching, and on the JUGENE machine of the JSC using the grand HHD19. This work was partially supported by the National Institute for Computational Sciences (NICS) under TG-MCA99S024 and utilized the NICS Kraken system. This collaboration was made possible through the support of the Annette Kade Graduate Student Fellowship Program at the American Museum of Natural History. NR also wants to thank IMPRS-HD.}

\appendix

\section{Numerical artefacts}
Shearing sheet simulations with the TRAMP code have displayed unreliable behavior for the extreme cases of cooling times, either isothermal ($\tau_{\rm cool}=0$) or
adiabatic ($\tau_{\rm cool}=\infty$). In the first case, a global pressure gradient in a locally isothermal disk leads to the amplification of radially propagating sound waves, which is a physically realistic case (see the derivation in Klahr 2013 ApJ submitted), but only shows up in local radially periodic simulations because the sound wave can propagate through the the box for an unlimited amount of time, which of course is not possible in a global disk. This physical instability can thus be found both in 1D radial TRAMP as well as in {\sc Pencil Code} simulations with remarkably identical growth behavior. This means, having a too short cooling time artifacts from these radially propagating sound waves could ruin our models. Nevertheless, as pointed out by Klahr (2013 ApJ submitted) already a cooling time of $\tau_{\rm cool}=0.01$ will suppress these sound wave instability completely.

\begin{figure}{!h}
\includegraphics[width=0.5\textwidth]{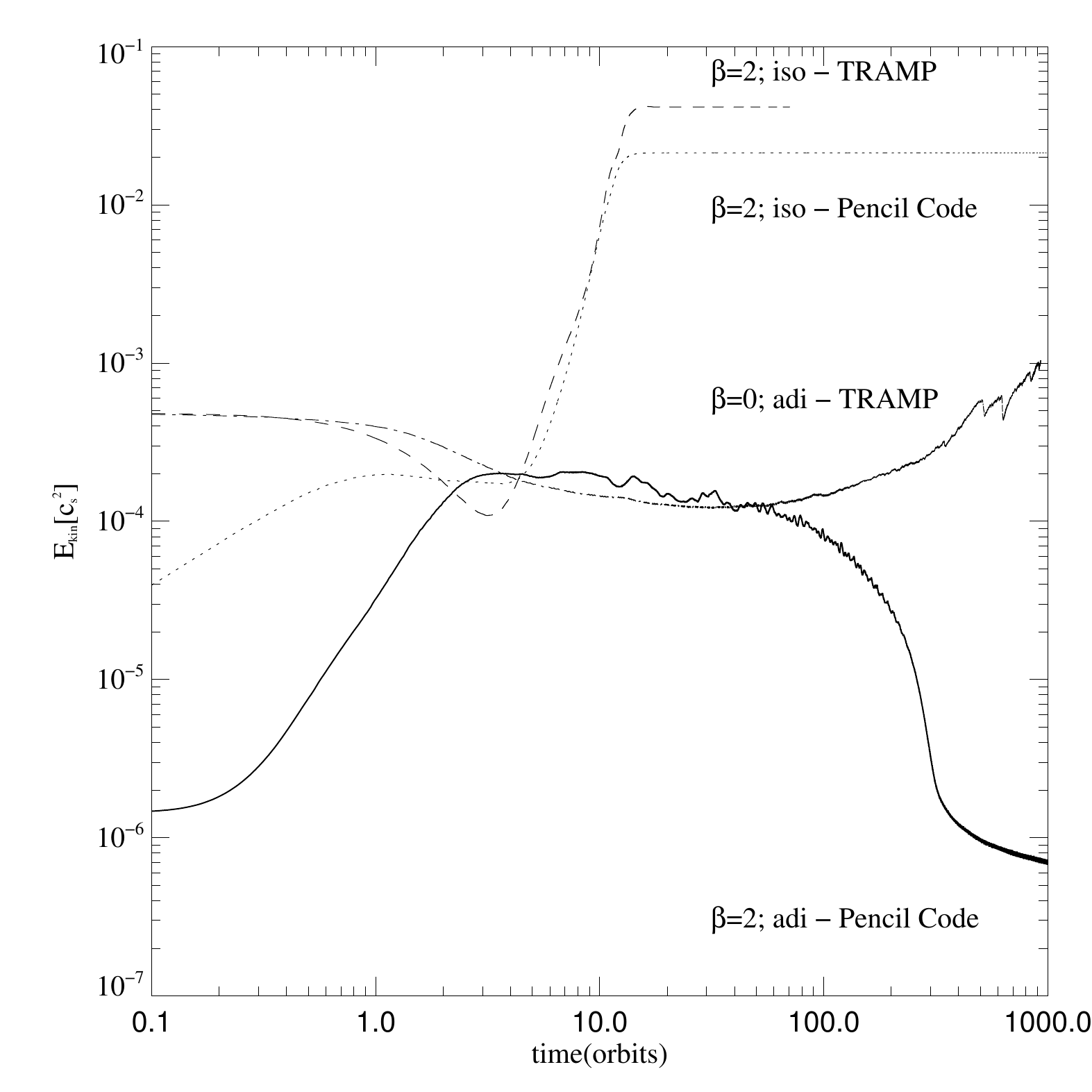}
\caption{Comparison of the kinetic energy for isothermal and adiabatic setup with the TRAMP code and the {\sc Pencil Code}. Both codes show the identical behavior for the isothermal case (dashed and dotted lines), yet in the adiabatic case the TRAMP code shows an artificial amplification of kinetic energy (dashed-dotted line). The {\sc Pencil Code} does not show this behavior.}
\label{error}
\end{figure}

On the other hand the adiabatic simulations using the TRAMP code were showing a weak amplification of kinetic energy over very long time scales which is the accumulation of numerical error in the quasi dissipation free TRAMP scheme. This behavior is independent of the chosen entropy gradient and results from the conservative treatment of Coriolis forces. Again the {\sc Pencil Code} with its explicit dissipation does not allow for this accumulation of this numerical error, even in the presence of a radial entropy gradient (see solid and dashed-dotted line in \Fig{error}).\\

\end{document}